\documentclass[aps,showpacs,pra,twocolumn]{revtex4}
\usepackage[pdftex]{graphicx}
\usepackage{amsmath}
\usepackage{amssymb}
\newcommand{\comment}[1]{}

\newcommand{\Esca}{\mathcal{E}}
\newcommand{\Bsca}{\mathcal{B}}
\newcommand{\Evec}{\vec{\mathcal{E}}}
\newcommand{\Bvec}{\vec{\mathcal{B}}}

\begin{document}

\title{Magnetic and electric dipole moments of the $H\ {}^3\Delta_1$ state in ThO}
\author{A. C. Vutha$^1$\email{amar.vutha@yale.edu}, B. Spaun$^2$, Y. V. Gurevich$^2$, N.R. Hutzler$^2$, E. Kirilov$^1$, J. M. Doyle$^2$, G. Gabrielse$^2$ and D. DeMille$^1$}
\affiliation{$^1$ Department of Physics, Yale University, New Haven, CT 06520, USA}
\affiliation{$^2$ Department of Physics, Harvard University, Cambridge, MA 02138, USA}

\begin{abstract}
The metastable $H \ {}^3\Delta_1$ state in the thorium monoxide (ThO) molecule is highly sensitive to the presence of a $CP$-violating permanent electric dipole moment of the electron (eEDM) \cite{MB08}. The magnetic dipole moment $\mu_H$ and the molecule-fixed electric dipole moment $D_H$ of this state are measured in preparation for a search for the eEDM. The small magnetic moment $\mu_H = 8.5(5) \times 10^{-3} \ \mu_B$ displays the predicted cancellation of spin and orbital contributions in a ${}^3 \Delta_1$ paramagnetic molecular state, providing a significant advantage for the suppression of magnetic field noise and related systematic effects in the eEDM search. In addition, the induced electric dipole moment is shown to be fully saturated in very modest electric fields ($<$ 10 V/cm). This feature is favorable for the suppression of many other potential systematic errors in the ThO eEDM search experiment. 
\end{abstract}

\maketitle

Measurable $CP$-violation in physics beyond the Standard Model is predicted in many proposed extensions to the Standard Model, and could provide a clue to the observed dominance of matter over antimatter in the universe \cite{FSB03}. The permanent electric dipole moment of the electron (eEDM) is a sensitive probe for flavor-diagonal $CP$-violation in the lepton sector \cite{KL97}. A number of experimental efforts are currently focused on searching for this elusive quantity \cite{RM10}. Many of these experiments take advantage of the large internal electric field $\Esca_{mol}$ experienced by valence electrons in a polar molecule \cite{San67}. Following the suggestion of Meyer \emph{et al.} \cite{MBD06}, states with a ${}^3 \Delta_1$ character in heavy molecules are being used in several new eEDM experiments \cite{LMP+09,VCG+10,LBL+10}. In addition to a large intrinsic eEDM sensitivity, there are two key attractive features of this kind of molecular state for eEDM searches: closely spaced opposite parity doublets ($\Omega$-doublets) \cite{DBB+01, MBD06} and small magnetic moments \cite{Cra34, MBD06}. The $\Omega$-doublets allow complete polarization of the molecule in small electric fields, taking full advantage of the eEDM sensitivity of the molecule while suppressing $\Esca$-field induced systematic errors, such as those due to leakage currents and geometric phases \cite{KBB+04,BHJD08,VCG+10}. The extremely small magnetic moment of the $H \ ^3\Delta_1$ state makes the molecule less sensitive to effects arising from fluctuating $\Bsca$-fields and motional $(\vec{v} \times \Evec/c^2)$ magnetic fields \cite{VCG+10}. Here we report measurements of both of these key parameters in the $H$ state of ThO. \comment{In addition to a large eEDM sensitivity, attractive features of these molecules such as combination of closely spaced opposite parity doublets  \cite{} and an extremely small magnetic moment in a \ ${}^3\Delta_1$ state \cite{Cra34} are convenient for suppressing many sources of noise and systematic errors \cite{MBD06,VCG+10}. Here we report on measurements of the small magnetic moment of the metastable $H$ state in thorium monoxide (ThO), and demonstrate the large polarizability of the $\Omega$-doublets in this state. These properties of the $H$ state are crucial to the effort to measure the eEDM using ThO . }

\section{Experimental setup}
The measurements were carried out using a molecular beam of ThO, produced in an apparatus similar to one described elsewhere \cite{VCG+10}. The apparatus uses helium buffer gas at 4 K to cool a pulse of ThO molecules (produced by pulsed laser ablation of ThO$_2$), which are extracted into a molecular beam and probed 30 cm downstream. The lowest rovibrational level ($v=0,J=1$) in the $H$ state was populated by optical pumping from the ground electronic $X \ {}^1\Sigma^+ (v=0,J=1) $ state via the higher-lying, short-lived $A \ {}^3\Pi_{0^+} (v=0,J=0)$ state. It was subsequently probed a few mm downstream by exciting laser-induced fluorescence (LIF). Both the 943 nm light to drive the $X \rightarrow A$ pump transition and the 908 nm light for the $H \rightarrow E$ probe transition were derived from external cavity diode lasers. Fluorescence from $E \rightsquigarrow X$ at 613 nm was collected with an f/1.0 lens, channeled through a quartz lightpipe and a bandpass interference filter and monitored with a photomultiplier tube. Detection of fluorescence at a wavelength significantly to the blue of the excitation laser suppresses background due to scattered laser light. The pump and probe lasers were perpendicular to the molecular beam and the transverse Doppler width on the $H \to E$ probe transition was $\sim$5 MHz. With the pump laser locked to resonance, the probe laser's frequency was tuned. The frequency steps were calibrated by monitoring the laser's transmission through a scanning confocal interferometer, which was actively stabilized to a 1064 nm YAG laser (in turn locked to an iodine cell) \cite{Far07}. The free spectral range of the interferometer was independently determined from its length, and from off-line measurements of spectra with RF sidebands added to the laser.
	
\section{Magnetic dipole moment}
A ${}^3\Delta_1$ molecular state has 2 units of orbital angular momentum ($\Lambda = 2$) and one unit of spin angular momentum ($\Sigma = -1$) projected onto the inter-nuclear axis. The contributions of these to the magnetic moment cancel out to a large extent, since the orbital $g$-factor ($g_L=1$) is very nearly half as large as the spin $g$-factor ($g_S = 2.002$); hence the effective $g$-factor is $g_{\textrm{eff}} = g_L \Lambda + g_S \Sigma \approx 0$ \cite{MBD06,Cra34}. The magnetic moment of such a pure molecular state is nonzero only because of small effects such as the nonzero value of $g_S-2$. \emph{Ab initio} calculations indicate that the $H$ state in ThO has 99\% ${}^3\Delta_1$ content, with small admixtures of other Hund's case (a) states (${}^1\Pi_1$, ${}^3\Pi_1$) due to off-diagonal spin-orbit mixings \cite{PNH+03}. These spin-orbit admixtures are expected to be the dominant contribution to the non-zero magnetic moment of the $H$ state, at the level of $g_{\textrm{eff}} \sim 0.01$. Other effects such as the magnetic moment due to the rotation of this polar molecule are expected to be much smaller \cite{Sau98}. 
\begin{figure}
\centering
\includegraphics[width=\columnwidth]{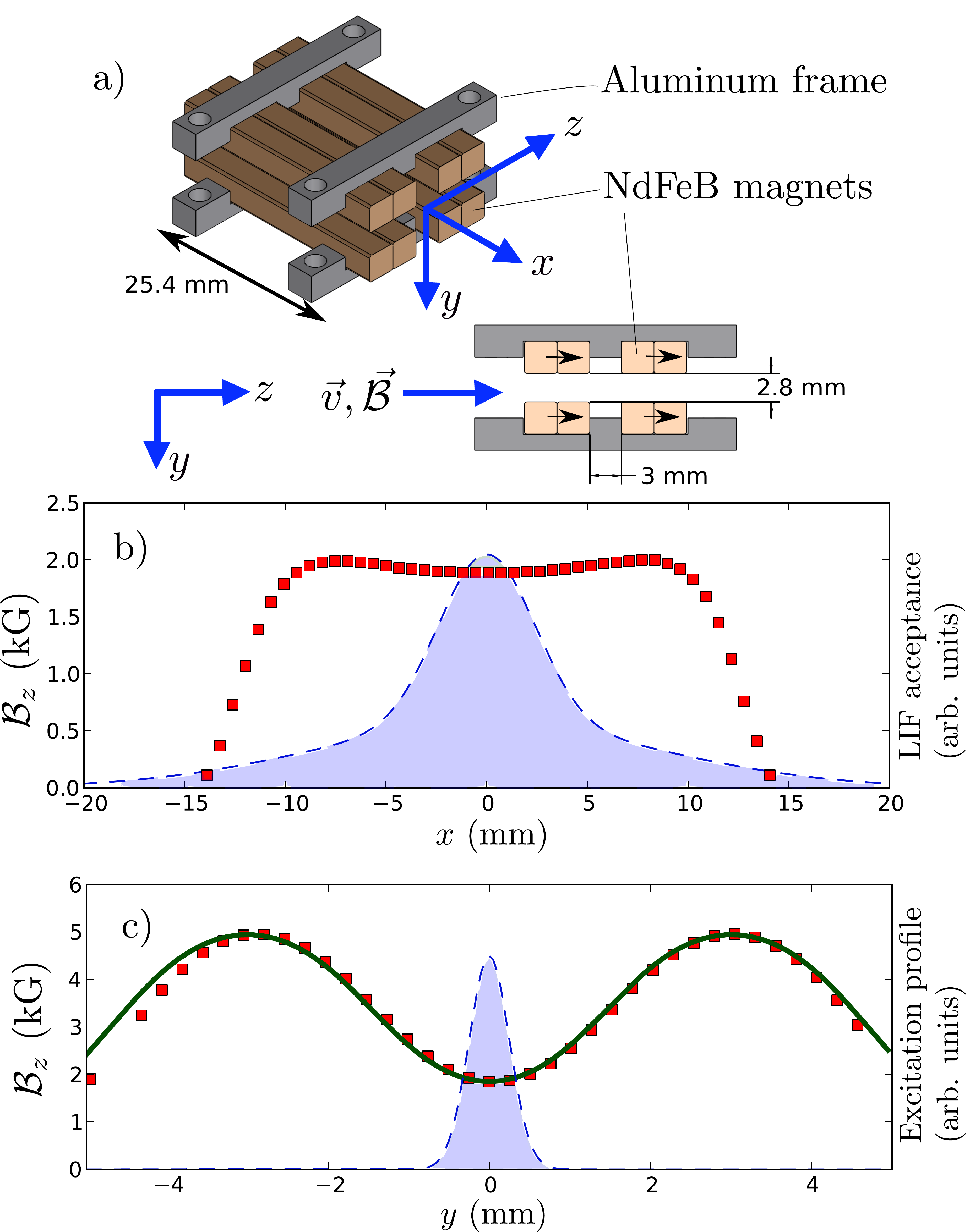}
\caption{a) The magnet assembly used in the measurement of $\mu_H$. Arrows on the NdFeB magnets indicate the direction of magnetization. b) The $\hat{z}$ component of the $\Bsca$-field, $\Bsca_z$, measured along the $x$-axis. The shaded area indicates the calculated acceptance of the LIF detection optics. c) Measured (red squares) and calculated (solid green line) values of $\Bsca_z$ along the $y$-axis. The shaded area indicates the LIF excitation region, defined by the probe laser's intensity profile.}
\label{fig:one}
\end{figure}

In order to split the $m_J$ azimuthal sublevels in the $H$ state by a frequency greater than the Doppler width in our molecular beam, it was found necessary to apply a large magnetic field $\mathcal{B} \geq 1$ kG. To do this, we constructed a compact permanent magnet assembly using NdFeB magnets (see Fig.\ref{fig:one}). The separation and alignment of the magnets was adjusted to obtain a uniform magnetic field over the region probed by the laser. The magnet was oriented so that $\Bvec \parallel \vec{v}$, the velocity of the molecular beam, in order to avoid spurious effects due to motional electric fields ($\Evec_{mot} = \vec{v} \times \Bvec$) polarizing the molecular state. The probe laser was collimated (intensity FWHM $\simeq$ 0.6 mm) to spatially select a well-defined region near the center of the magnet assembly; its $k$-vector was aligned along $\hat{x}$ in Fig.\ref{fig:one}a. The polarization $\hat{\epsilon}$ of the probe laser was adjusted to be parallel (perpendicular) to the $\Bsca$-field in order to probe the unshifted $m_J=0$ (Zeeman-shifted $m_J=\pm1$) states. The LIF collection lens and optics were positioned along $\hat{y}$.

\begin{figure}
\centering
\includegraphics[width=\columnwidth]{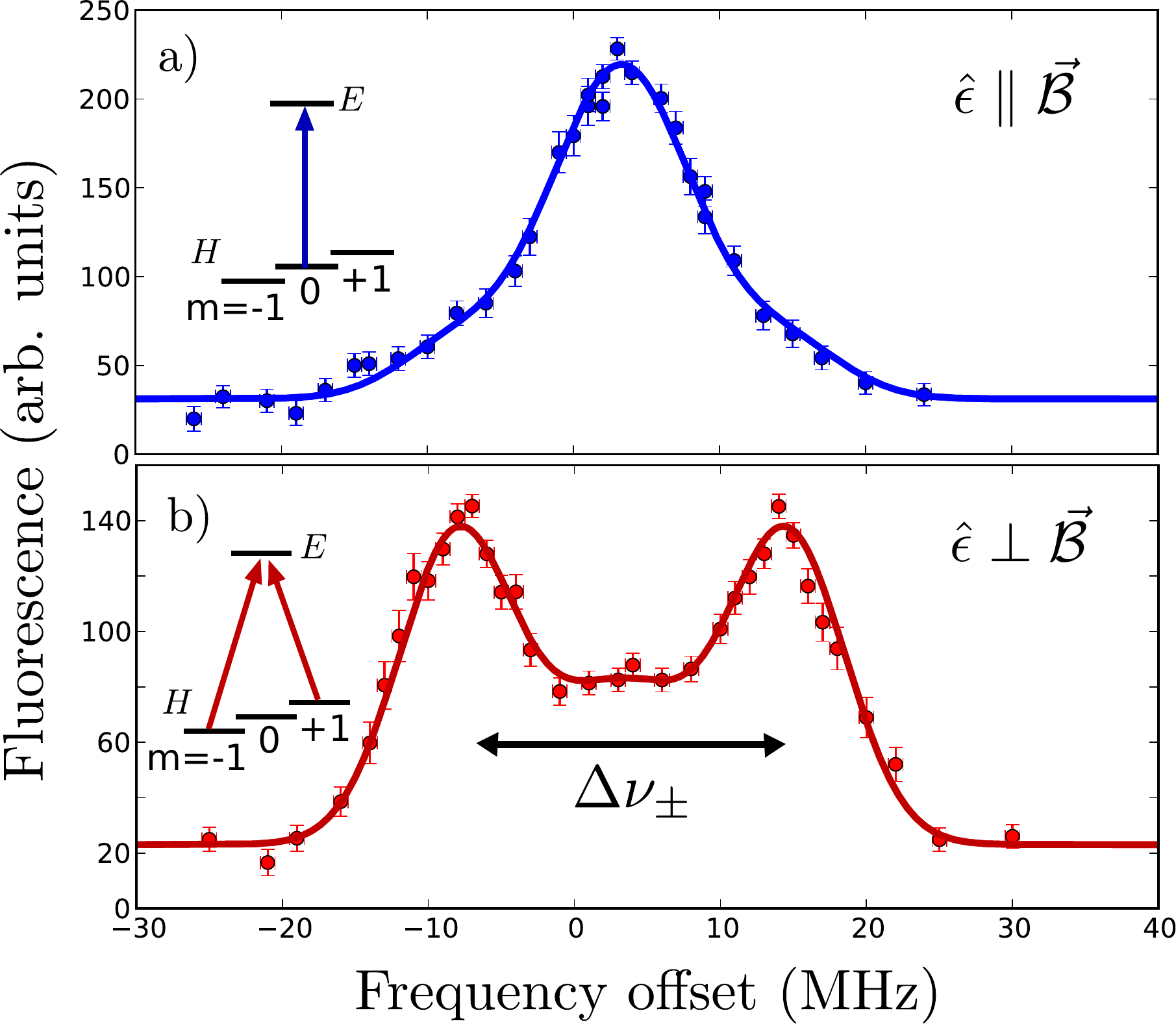}
\caption{Spectra of LIF from the $H,J=1$ state in a magnetic field $\Bsca$ = 1.9(1) kG, acquired with 16 averages per data point. The $x$-error bars account for the standard error in the laser's frequency offset (derived from the rms frequency excursion of the lock's error signal) and $y$-error bars indicate the standard error of the LIF signal due to shot-to-shot fluctuations in the yield of molecules in a pulse. a) The probe laser's polarization $\hat{\epsilon} \parallel \Bvec$; in this configuration the $m_J = 0$ sublevel is probed. b) With the laser polarization $\hat{\epsilon} \perp \Bvec$, the $m_J = \pm1$ sublevels are probed. The fit to the spectrum in (b) yields $\Delta \nu_\pm$ = 22.66(41) MHz for the Zeeman splitting between $m_J=\pm1$.}
\label{fig:two}
\end{figure}

The spectra of LIF collected from the $H, J=1$ state in the magnetic field region are shown in Fig. \ref{fig:two}. Spectra with $\hat{\epsilon} \parallel \Bvec$ and $\hat{\epsilon} \perp \Bvec$ were simultaneously fit to a sum of 3 gaussian lineshapes, with the line centers and linewidths constrained to have the same value for both data sets. We included `minority' peaks corresponding to the orthogonal polarization (i.e. the $m_J=0$ peak in the $\hat{\epsilon} \perp \Bvec$ fit, and the $m_J = \pm1$ peaks in the $\hat{\epsilon} \parallel \Bvec$ fit) to model the effect of residual circular polarization (due to an imperfect waveplate and birefringence in vacuum windows). The measured amount of circular polarization ($\approx$ 15\%) was in fair agreement with that deduced from the relative size of the minority peaks ($\approx$ 20\%). We verified that changes in the size of the minority peaks did not affect the Zeeman shift extracted from the fit within its uncertainty. Varying the width of the minority peaks shifted the fit value of $\Delta_\pm$ by 0.33 MHz; this effect is included as a systematic contribution to the fit uncertainty. The fit frequency separation between the $m_J=\pm1$ peaks in Fig.1b is $\Delta \nu_\pm$ = 22.66(41) MHz. $\Delta \nu_\pm$ is related to the intrinsic magnetic moment of the H state, $\mu_H$, by the formula $h \Delta \nu_\pm = \frac{2 \mu_H \Bsca}{J(J+1)}$ \cite{BC03}. \comment{Analysis of the data from 2010/10/06 yields: $\Delta \nu \pm$ = 22.70(25,stat) MHz.}

The magnetic field sampled by molecules in the experiment was characterized as follows. The magnetic field profiles in Fig.\ref{fig:one}b,c were measured with a Hall probe whose active area (0.127 mm $\times$ 0.127 mm) is small compared to the area illuminated by the laser beam. Spatial selectivity along the $x$-axis was provided by the LIF detection optics, as shown in Fig.\ref{fig:one}b. The acceptance of the optical system was modeled with ray-tracing software and used to extract the central value of the $\Bsca$-field along this dimension. As the separation between the magnets was too small to allow a direct measurement of $\Bsca_z$ along $\hat{z}$, we accounted for the spatial dependence in the $yz$ plane in the following way. The value of the $\Bsca$-field measured at the origin in the $\Bsca_z$ vs. $y$ profile was used to calibrate the pole strength in a 2D numerical calculation in the $yz$ plane based on the (measured) magnet geometry. When weighted over an area corresponding to the Hall probe, the calculation reproduced the measured $\Bsca$-field profile, as shown in Fig.\ref{fig:one}c. The calculated $\Bsca$-field pattern was weighted by the Gaussian intensity profile of the probe laser beam in the $yz$ plane and averaged. (The LIF yield on the $H \to E$ detection transition was linear in the probe laser's intensity). We calculated the average $\Bsca$-field, $\langle \Bsca \rangle$, in this way for a range of displacements of the laser beam profile ($\pm$ 0.5 mm) from the exact center of the magnets in the $yz$ plane to account for the experimental uncertainty in the laser's position and pointing. This leads to the estimate that $\langle \Bsca \rangle$ = 1.9(1) kG in the probed volume. As a further check on systematic errors arising from the probe laser's alignment, we repeated the measurement with a complete realignment of the laser's path through the magnet and obtained a value for $\Delta_\pm$ that was within the estimated range of possible changes in $\langle \Bsca \rangle$ due to misalignment (5\%).\comment{($\Delta_\pm$ = 22.70(40) MHz)} 

We combine the fit uncertainty in quadrature with uncertainties in the probe laser's frequency calibration (1\%) and the central value of the $\Bsca$-field in the probed volume (5\%) to obtain $|\mu_H|$ = 0.0085(5) $\mu_B$ for the magnetic moment of the $H$ state. 

\section{Molecule-fixed electric dipole moment}
The presence of $\Omega$-doublets, levels of opposite parity spaced much closer than the rotational splittings, in a ${}^3\Delta_1$ state leads to a polarizability that is typically $\geq 10^9$ atomic units. This means that the molecule can be fully polarized (i.e. the levels of opposite parity completely mixed) even in a static electric field as small as a few V/cm. This feature enables the suppression of a number of systematic errors in the measurement of the eEDM (see \cite{VCG+10} for more details). We measured the molecule-fixed dipole moment $D_H$ of the $H$ state by spectroscopically resolving the Stark shift in an applied electric field, and verified that the molecule was completely polarized in electric fields as low as 10 V/cm. 
\begin{figure}
\centering
\includegraphics[width=\columnwidth]{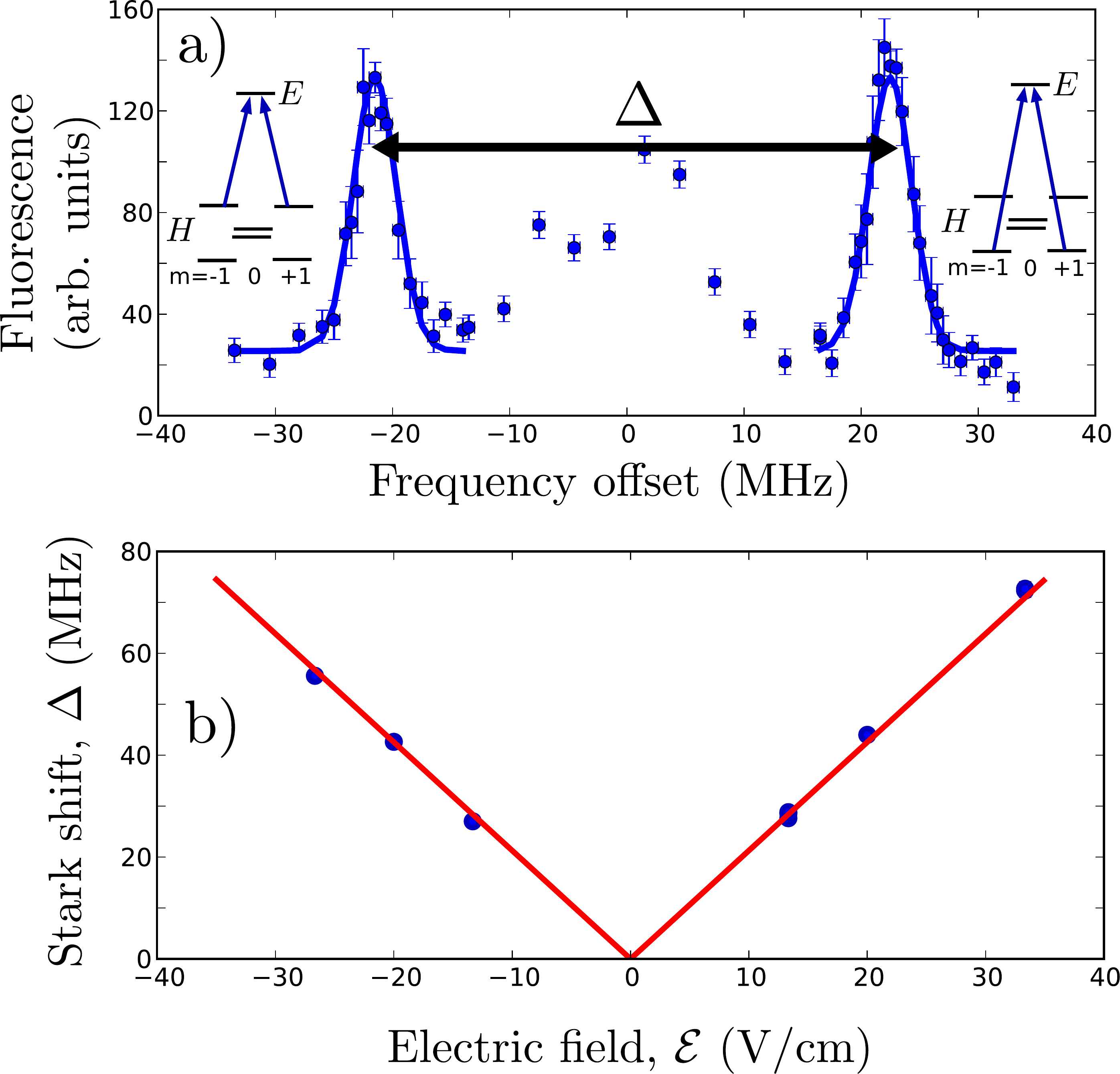}
\caption{a) LIF spectrum from the $H,J=1$ state in an electric field $\Esca = $20 V/cm, acquired with 16 averages per data point. The error bars are assigned in the same way as in Figure \ref{fig:two}. The two $\Esca$-field polarized states are separated by a frequency $\Delta$. The LIF signal near zero offset is due to molecules outside the electric field plates (see text). In b), $\Delta$ is plotted as a function of the $\Esca$-field across the plates. \comment{The $x$-error bars account for the uncertainty in the voltage applied to the plates, and the $y$-error bars are the fit uncertainty in the splitting between the $\Esca$-field polarized states.} The solid line is a fit to the function $\Delta = |D_J \Esca|$. The slope of the plot yields $D_{J=1} = h \times 2.13(2)$ MHz/(V/cm) for the dipole matrix element between the $\Omega$-doublets in $H, v=0, J=1$ state.}
\label{fig:three}
\end{figure}

For these measurements, a pair of glass plates, coated with transparent conducting indium tin oxide (ITO) on one side and broadband anti-reflection coating on the other, were used to make a capacitor with 25 mm x 30 mm plates separated by a 3.00(5) mm vacuum gap. The molecular beam was passed between the plates and a linearly polarized probe laser was sent at right angles to the molecular beam, normal to and through the transparent plates. LIF was collected with the same optical arrangement used for the magnetic moment measurement, at right angles to both the molecular beam velocity and the probe laser's $k$-vector. 

In the $H (v=0)$ state, we focus our analysis on the $J^p=1^\pm,|m_J|=1$ $\Omega$-doublet states ($p$ denotes the parity of the state). In the absence of an electric field, these states are parity eigenstates and are separated by an energy $\Delta_0$. (The Zeeman sublevels with $m_J=0$ do not mix in an electric field as a result of angular momentum selection rules, and we ignore them in the rest of this analysis.) In the two-state space spanned by the basis states $J=1^\pm,m_J=+1$ (or identically, the space with $m_J = -1$, since $m_J=\pm1$ are degenerate in an $\Esca$-field), the system in an $\Esca$-field is described by the Hamiltonian $H = \Big(\begin{array}{cc} -\Delta_0/2 & -D_J \Esca \\ -D_J \Esca & \Delta_0/2 \end{array}\Big)$, where $D_J$ is the electric dipole matrix element connecting the two basis states. The energy spacing between the eigenstates is $\Delta = 2 \sqrt{\Delta_0^2/4 + D_J^2 \Esca^2}$. In the limit where $|D_J \Esca| \gg \Delta_0$, the parity eigenstates are completely mixed and the energy spacing between the polarized eigenstates is $\Delta \approx 2 |D_J \Esca|$. 

In our experiment, the probe laser couples the $\Esca$-field-polarized states in $H (v=0)$ to the $E (v=0,J^p=0^+)$ state for LIF. The excited $E$ state does not have $\Omega$-doublets, and in an $\Esca$-field the predominant mixing of the $E,J^p=0^+$ state is with the neighboring $E,J^p=1^-$ rotational state. Since the rotational spacing in the $E$ state ($\sim$ 20 GHz) is much larger than the zero-field $\Omega$-doublet spacing ($\Delta_0 \sim$ 400 kHz) in the $H$ state \cite{EL84}, whereas the dipole matrix elements are of comparable size, there is a range of electric fields (1 V/cm $\lesssim |\Esca| \lesssim$ 1 kV/cm) where the $H, J^p=1^\pm$ $\Omega$-doublets are fully mixed while the $E,J^p=0^+$ state remains a parity eigenstate to a good approximation. In this regime therefore, LIF signals from both the polarized eigenstates in $H$ should be visible with equal intensity. Since the laser polarization $\hat{\epsilon} \perp \Evec$, only $|m_J|=1$ states are excited by the laser.

The sample LIF spectrum in Fig.\ref{fig:three}a shows peaks from the $\Esca$-field polarized eigenstates that are separated by a frequency $\Delta$. They are also well separated from the residual $\Esca$-field-free signal, which was due to molecules excited outside the capacitor plates. The narrowness of the $\Esca$-field polarized spectral peaks was due to additional collimation of the molecular beam by the capacitor plates. A pair of identical Gaussian lineshapes were fit to the $\Esca$-field polarized spectral peaks and their separation $\Delta$ was extracted. The $\Esca$-field-free signal was ignored for the purpose of fitting and extracting $\Delta$ from the data (we verified that the residual slope due to this signal did not affect the fit value of $\Delta$ within its uncertainty). In Fig.\ref{fig:three}b, the frequency separations $\Delta$ extracted from a set of LIF spectra are shown plotted against the $\Esca$-field across the plates. The linear dependence of $\Delta$ as a function of $|\Esca|$ indicates that the $H$ state was fully polarized over the range of $\Esca$-fields applied during the measurements. The fit yields the value $D_{J=1} = h \ \times $ 2.13(2) MHz/(V/cm), and constrains $\Delta_0 \leq$ 2 MHz in agreement with the result of \cite{EL84}. The relation between $D_J$ and the molecule-fixed dipole moment in the $H$ state, $D_H$, is $D_J = \frac{D_H}{J(J+1)}$ \cite{BC03}. We find $|D_H| = $ 1.67(4) $e a_0$, where the reported error is a quadrature sum of the fit uncertainty (1\%), and systematic uncertainties due to laser frequency calibration (1\%) and field plate spacing (2\%). 

\section{Summary}
We have measured the magnetic moment $\mu_H$ and molecule-fixed electric dipole moment $D_H$ of the metastable $H$ state in the ThO molecule. The suppression of $\mu_H$ predicted for an eEDM-sensitive ${}^3\Delta_1$ molecular state has been experimentally verified. The $H$ state in ThO can be polarized with very small electric fields due to the presence of $\Omega$-doublets. This combination of a small magnetic moment and large polarizability in the $H$ state enables the strong suppression of systematic effects in our ongoing experiment to search for the eEDM with ThO.

A.V. acknowledges helpful discussions with Alexei Buchachenko. We thank Elizabeth Petrik and Paul Hess for technical assistance and Wes Campbell for comments on the manuscript. This work was supported by the National Science Foundation.

\bibliography{hstate-moments}

\end{document}